\documentclass{llncs}

\include{diagrams}

\usepackage{oldlfont,amssymb,epsf,stmaryrd}
\usepackage{epsfig}

\newbox\tempa
\newbox\tempb
\newdimen\tempc
\def\mud#1{\hfil $\displaystyle{\mathstrut #1}$\hfil}
\def\rig#1{\hfil $\displaystyle{#1}$}
\def\irulehelp#1#2#3{\setbox\tempa=\hbox{$\displaystyle{\mathstrut #2}$}%
                        \setbox\tempb=\vbox{\halign{##\cr
        \mud{#1}\cr
        \noalign{\vskip\the\lineskip}%
        \noalign{\hrule height 0pt}%
        \rig{\vbox to 0pt{\vss\hbox to 0pt{${\; #3}$\hss}\vss}}\cr
        \noalign{\hrule}%
        \noalign{\vskip\the\lineskip}%

        \mud{\copy\tempa}\cr}}%
                      \tempc=\wd\tempb
                      \advance\tempc by \wd\tempa
                      \divide\tempc by 2 }
\def\irule#1#2#3{{\irulehelp{#1}{#2}{#3}%
                     \hbox to \wd\tempa{\hss \box\tempb \hss}}}

\begin{document}

\title{Cut-elimination and the decidability of reachability 
in alternating pushdown systems}
\author{Gilles Dowek\inst{1} \and Ying Jiang\inst{2}}
\institute{Inria, 
23 avenue d'Italie,
CS 81321, 75214 Paris Cedex 13, France, 
{\tt gilles.dowek@inria.fr}.
\and
State Key Laboratory of Computer Science,
Institute of Software, 
Chinese Academy of Sciences,
P.O. Box 8718, 100190 Beijing, China,
{\tt jy@ios.ac.cn}.}

\date{}
\maketitle
\thispagestyle{empty}

\begin{abstract}
We give a new proof of the decidability of reachability in alternating
pushdown systems, showing that it is a simple consequence of a 
cut-elimination theorem for some 
natural-deduction style inference systems. Then, we show how this result
can be used to extend an alternating pushdown system into a complete
system where for every configuration $A$, either $A$ or $\neg A$ is 
provable.
\end{abstract}

\section{Introduction}

Several methods can be used to prove that a problem is decidable. 
One of them is to reduce this problem to provability in some 
logic and prove that provability in this logic is decidable. Another is to 
reduce this problem to reachability in some transition system
and prove that reachability is decidable in this transition 
system.

For instance deciding if a number $n$ is even can be reduced to deciding
if the proposition $even(S^n(0))$ is provable in 
the logic defined by the rules 
$$\irule{}
        {even(0)}
        {}$$
$$\irule{even(x)} 
        {odd(S(x))}
        {}$$
$$\irule{odd(x)}
        {even(S(x))}
        {}$$
It can also be reduced to decide if the configuration $f$ is reachable 
from the configuration $\langle even, S^n0 \rangle$ in the pushdown system 
$$\langle even, 0 \rangle \hookrightarrow f$$
$$\langle even, S w \rangle \hookrightarrow \langle odd, w \rangle$$
$$\langle odd, S w \rangle \hookrightarrow \langle even, w \rangle$$

Although at a first glance, logics and transition systems look alike
as they both define a set of {\em things}---propositions, states,
configurations---and {\em rules}---deduction rules, transition rules---to go
step by step from one thing to another, the details look quite
different.  In particular, the methods used to prove the decidability
of provability in 
a logic---quantifier-elimination, finite model property,
cut-elimination, etc.---and those used to prove the decidability of
reachability in a transition system---finite state automata,
etc.---are not easy to relate.

In this paper, we establish a connection between
proof-theoretical methods and automata-theoretical methods to prove the
decidability of a problem. In particular we show that the run of 
an automaton can be seen as a cut-free proof and the proof that 
the set of reachable configurations in a transition system can be 
recognized by a finite-state automaton as a cut-elimination theorem.

More precisely, in Section \ref{secdec}, we prove a cut-elimination
theorem for a class of logics and show that the decidability of
reachability in alternating pushdown systems is a consequence of this
cut-elimination theorem. The decidability of reachability in
alternating pushdown systems \cite{BEM}, is a seminal result in
automata theory as many other results, such as the decidability of
LTL, CTL, and the $\mu$-calculus over pushdown systems, are
corollaries.  In Sections \ref{seccoco} and \ref{seccoindind}, 
we relate the notion of negation as
failure and of complementation of an automaton, and prove how this
decidability result permits to design a complete logic, where for
each closed proposition, either $A$ or $\neg A$ is provable.

\section{Decidability} 
\label{secdec}

In this section, we define a class of logics, called {\em alternating 
pushdown systems} and prove the decidability of provability in these logics.

\begin{definition}[State, word, configuration]
Consider a language ${\cal L}$ in monadic predicate logic, containing a finite
number of predicate symbols, called {\em states}, a finite
number of function symbols, called {\em stack symbols}, and a
constant $\varepsilon$, called {\em the empty word}.

A closed term in ${\cal L}$ has the form 
$\gamma_1 (\gamma_2 ... (\gamma_n (\varepsilon)))$ where $\gamma_1$,
..., $\gamma_n$ are stack symbols.  Such a term is
called a {\em word} and is often written $w = \gamma_1 \gamma_2
... \gamma_n$. An open term has the form $\gamma_1 (\gamma_2
... (\gamma_n (x)))$ for some variable $x$. It is often written
$\gamma_1 \gamma_2 ... \gamma_n x$ or $wx$ for $w = \gamma_1 \gamma_2
... \gamma_n$. 

A closed atomic proposition, called a {\em configuration}, has the
form $P(w)$ where $P$ is a state and $w$ a word. An open atomic
proposition has the form $P(w x)$ where $P$ is a state, $w$ a word,
and $x$ a variable.
\end{definition}

\begin{definition}[Alternating pushdown system]
An {\em alternating pushdown system} is given by a finite set of
inference rules, called {\em transition rules}, of the form
$$\irule{P_1(v_1 x)~...~P_n(v_n x)}{Q(w x)}{}$$
where $v_1, ..., v_n, w$ are words and $n$ may be zero,
or of the form
$$\irule{}
        {Q(\varepsilon)}
        {}$$
\end{definition}

A rule of the first form may also be written as
$$\langle Q, w x \rangle \hookrightarrow 
\{\langle P_1, v_1 x \rangle, ..., \langle P_n, v_n x \rangle\}$$
or simply
$$\langle Q, w \rangle \hookrightarrow 
\{\langle P_1, v_1 \rangle, ..., \langle P_n, v_n \rangle\}$$
and a rule of the second form may also be written as
$$\langle Q, \varepsilon \rangle \hookrightarrow \varnothing$$

\begin{definition}[Proof]
\label{proof}
A {\em proof} in an inference system ${\cal I}$ 
is a finite tree labeled by configurations such that 
for each node $N$, there exists an inference rule 
$$\irule{A_1~...~A_n}{B}{}$$
in ${\cal I}$, 
and a substitution $\sigma$ such that the node $N$ is labeled with 
$\sigma B$ and its children are labeled with
$\sigma A_1$, ..., $\sigma A_n$. 

A proof is a {\em proof of a configuration} $A$ if 
its root is labeled by $A$.

A configuration $A$ is said to be {\em provable}, written 
$A \in pre^*(\varnothing)$, if it has a proof.
\end{definition}

\begin{example}
\label{ex}
In the system 
$$\begin{array}{llll}
\irule{Q(x)}
      {P(a x)}
      {\mbox{\bf i1}}
~~~~~~~~~~~~~~
&
\irule{T(x)}
      {P(b x)}
      {\mbox{\bf i2}}
~~~~~~~~~~~~~~
&
\irule{T(x)}
      {R(a x)}
      {\mbox{\bf i3}}
~~~~~~~~~~~~~~
&
\irule{}
      {R(b x)}
      {\mbox{\bf i4}}
\\
\\
\irule{P(x)~R(x)}
      {Q(x)}
      {\mbox{\bf n1}}
~~~~~~~~~~~~~~
&
\irule{}
      {T(x)}
      {\mbox{\bf n2}}
\\
\\
\irule{P(ax)}
      {S(x)}
      {\mbox{\bf e1}}
\end{array}$$
the configuration $S(ab)$ has the following proof
$$\irule{\irule{\irule{\irule{\irule{\irule{\irule{}{T(\varepsilon)}{\mbox{\bf n2}}}
                                           {P(b)}
                                           {\mbox{\bf i2}}
                                     ~~~~~~~~~~~~~~~
                                     \irule{}
                                           {R(b)}
                                           {\mbox{\bf i4}}
                                    }
                                    {Q(b)}
                                    {\mbox{\bf n1}}
                             }
                             {P(ab)}
                             {\mbox{\bf i1}}
                      ~~~~~~~~~~~~~~~~~~~~~~~
                      \irule{\irule{}{T(b)}{\mbox{\bf n2}}}
                            {R(ab)}
                            {\mbox{\bf i3}}
                      }
                      {Q(ab)}
                      {\mbox{\bf n1}}
               }
               {P(aab)}
               {\mbox{\bf i1}}
         }
         {S(ab)}
         {\mbox{\bf e1}}$$
\end{example}

This proof can also be written 
$\{S(ab)\} 
\hookrightarrow 
\{P(aab)\}
\hookrightarrow 
\{Q(ab)\}
\hookrightarrow 
\{P(ab),R(ab)\}
\hookrightarrow 
\{Q(b),R(ab)\}
\hookrightarrow 
\{Q(b),T(b)\}
\hookrightarrow 
\{Q(b)\}
\hookrightarrow 
\{P(b),R(b)\}
\hookrightarrow 
\{P(b)\}
\hookrightarrow 
\{T(\varepsilon)\}
\hookrightarrow 
\varnothing$.

\begin{definition}[Introduction rule, elimination rule, neutral rule] 
An {\em introduction rule} is a rule of the form 
$$\irule{P_1(x)~...~P_n(x)}
        {Q(\gamma x)}
        {}$$
where $\gamma$ is a stack symbol, $n$ may be zero, or of the form
$$\irule{}
        {Q(\varepsilon)}
        {}$$
An {\em elimination rule} is a rule of the form 
$$\irule{P_1(\gamma x)~P_2(x)~...~P_n(x)}
        {Q(x)}
        {}$$
where $\gamma$ is a stack symbol and $n$ is at least one.\\
A {\em neutral rule} is a rule of the form
$$\irule{P_1(x)~...~P_n(x)} 
        {Q(x)} 
        {}$$ 
where $n$ may be zero.
\end{definition}

\begin{definition}[Alternating multi-automaton]
An alternating pushdown system of which all rules are introduction rules 
is called an {\em alternating multi-automaton}. If the configuration 
$P(w)$ is provable in an alternating multi-automaton, we say also that 
the word $w$ is {\em recognized in} $P$.
\end{definition}

The introduction rule
$$\irule{P_1(x)~...~P_n(x)}
        {Q(\gamma x)}
        {}$$
may be written as
$$\langle Q, \gamma x \rangle \hookrightarrow 
\{\langle P_1, x \rangle, ..., \langle P_n, x \rangle\}$$
or simply
$$\langle Q, \gamma \rangle \hookrightarrow 
\{\langle P_1, \varepsilon \rangle, ..., \langle P_n, \varepsilon \rangle\}$$
It is also sometime written as
$$Q \hookrightarrow^{\gamma} \{P_1, ..., P_n\}$$

\begin{lemma}[Decidability]
Provability is decidable in an alternating multi-automaton.
\end{lemma}

\proof{Bottom-up proof-search terminates as the size of configurations
decreases at each step.}

\medskip

If decidability is obvious for alternating multi-automata, it is 
less obvious for 
general alternating pushdown systems, 
as bottom-up proof-search, that is
eager application of the transition rules, does not always terminate,
even if we include a redundancy check \`a la Kleene \cite{Kleene}. For
instance, consider an alternating pushdown system containing the 
elimination rule
$$\irule{P(a x)}
        {P(x)}
        {}$$
applying this rule bottom-up to the configuration $P(a)$ yields
$P(aa)$, $P(aaa)$, $P(aaaa)$, ...

To prove the decidability of provability in arbitrary alternating 
pushdown systems, we shall prove a cut-elimination result and a
subformula property that permit to avoid considering configurations such 
as $P(aa)$, $P(aaa)$, etc., which are not subformulae of $P(a)$.

We start with a simple lemma, that permits to restrict to particular
alternating pushdown systems called {\em small step alternating 
pushdown systems}.

\begin{definition}[Small step alternating pushdown system]
A {\em small step} alternating pushdown system is an alternating
pushdown system of which each rule is either an introduction rule, an
elimination rule or a neutral rule.
\end{definition}

\begin{lemma}
For each alternating pushdown system ${\cal I}_0$, there exists a small step 
alternating pushdown system ${\cal I}$ that is a conservative extension 
of ${\cal I}_0$. 
\end{lemma}

\proof{Assume the system ${\cal I}_0$ contains a rule $r$ that is neither an 
introduction rule, nor an elimination rule, nor a neutral rule. 

For all propositions of the form $P(\gamma_1 ... \gamma_n x)$
occurring as a premise or a conclusion of this rule, we introduce $n$ 
predicate symbols 
$P^{\gamma_1}$,  $P^{\gamma_1 \gamma_2}$, ..., $P^{\gamma_1 ... \gamma_n}$,  
$n$ introduction rules 
$$\irule{P^{\gamma_1 ... \gamma_i \gamma_{i+1}}(x)}
        {P^{\gamma_1 ... \gamma_i }(\gamma_{i+1} x)}
        {}$$
and $n$ elimination rules 
$$\irule{P^{\gamma_1 ... \gamma_i }(\gamma_{i+1} x)}
        {P^{\gamma_1 ... \gamma_i \gamma_{i+1}}(x)}
        {}$$  
and we replace the rule $r$ by the neutral rule $r'$ obtained by 
replacing the proposition
$P(\gamma_1 ... \gamma_n x)$ 
by $P^{\gamma_1 ... \gamma_n}(x)$.

Obviously, this system is an extension of ${\cal I}_0$, as the rule $r$ is 
derivable from the rule $r'$ and the added introduction and elimination 
rules. And this extension is 
conservative as, by replacing the configuration $P^{\gamma_1 ... \gamma_i}(w)$ 
by $P(\gamma_1 ... \gamma_i w)$, we obtain a proof in the original system.

\begin{definition}[Cut]
A {\em cut} is a proof of the form
$$\irule{\irule{\irule{\pi_1}
                      {P_1(w)}
                      {}
                ~~~
                ...
                ~~~
                \irule{\pi_m}
                      {P_m(w)}
                      {}
               }
               {Q_1(\gamma w)}
               {\mbox{intro}}
               ~~~~~~~~~~~~~~~~~
               \irule{\rho_2}{Q_2(w)}{}
               ~...~
               \irule{\rho_n}{Q_n(w)}{}
        }
        {R(w)}
        {\mbox{elim}}
$$
$$\irule{\irule{\irule{\pi^1_1}
                      {P^1_1(w)}
                      {}
                ~...~
                \irule{\pi^1_{m_1}}
                      {P^1_{m_1}(w)}
                      {}
               }
               {Q_1(\gamma w)}
               {\mbox{intro}}
         ~~~~~~~~~~~~~~~~~~~
         ... 
         ~~~~~~~~~
         \irule{\irule{\pi^n_1}
                      {P^n_1(w)}
                      {}
                ~...~
                \irule{\pi^n_{m_n}}
                      {P^n_{m_n}(w)}
                      {}
                }
                {Q_n(\gamma w)}
               {\mbox{intro}}
        }
        {R(\gamma w)}
        {\mbox{neutral}}
$$
or 
$$\irule{\irule{}
               {Q_1(\varepsilon)}
               {\mbox{intro}}
         ~~~~~~~~
         ... 
         ~~
         \irule{}
               {Q_n(\varepsilon)}
               {\mbox{intro}}
        }
        {R(\varepsilon)}
        {\mbox{neutral}}
$$

A proof {\em contains a cut} if one of its subproofs is a cut. 
A proof is {\em cut-free} if it contains no cut.
A small step alternating pushdown system 
{\em has the cut-elimination property} if 
every provable configuration has a cut-free proof.
\end{definition}

Not all small step alternating pushdown systems have the
cut-elimination property.  For instance, in the system defined in
Example \ref{ex}, the configuration $S(ab)$ has a proof but no cut-free
proof.  Thus, instead of proving that every small step alternating
pushdown system has the cut-elimination property, we shall prove that
every small step alternating pushdown system has an extension with
derivable rules, that has the cut-elimination property.

Note the similarity between this method and the Knuth-Bendix method
\cite{KB}, which does not prove that all rewrite systems are confluent,
but instead that, in some cases, it is possible to extend a rewrite system
with derivable rules to make it confluent \cite{DershowitzKirchner}.

\begin{definition}[Saturation]
Consider a small step alternating pushdown system.
\begin{itemize}
\item
If the system contains an introduction rule
$$\irule{P_1(x)~...~P_m(x)}{Q_1(\gamma x)}{\mbox{intro}}$$
and an elimination rule
$$\irule{Q_1(\gamma x)~Q_2(x)~...~Q_n(x)}{R(x)}{\mbox{elim}}$$
then we add to it the neutral rule 
\smallskip
$$\irule{P_1(x)~...~P_m(x)~Q_2(x)~...~Q_n(x)}{R(x)}{\mbox{neutral}}$$
\item
If the system contains introduction rules 
$$\irule{P^1_1(x)~...~P^1_{m_1}(x)}{Q_1(\gamma x)}{\mbox{intro}}$$
$$...$$
$$\irule{P^n_1(x)~...~P^n_{m_n}(x)}{Q_n(\gamma x)}{\mbox{intro}}$$
and a neutral rule 
$$\irule{Q_1(x)~...~Q_n(x)}{R(x)}{\mbox{neutral}}$$
then we add to it the introduction rule 
$$\irule{P^1_1(x)~...~P^1_{m_1}(x)~...~P^n_1(x)~...~P^n_{m_n}(x)}
        {R(\gamma x)}
        {\mbox{intro}}$$
In particular, if the system contains a neutral rule 
$$\irule{}{R(x)}{\mbox{neutral}}$$
then we add to it the introduction rule 
$$\irule{}{R(\gamma x)}{\mbox{intro}}$$
for all $\gamma$.
\item
If the system contains introduction rules
$$\irule{}{Q_1(\varepsilon)}{\mbox{intro}}$$
$$...$$
$$\irule{}{Q_n(\varepsilon)}{\mbox{intro}}$$
and a neutral rule 
$$\irule{Q_1(x)~...~Q_n(x)}{R(x)}{\mbox{neutral}}$$
then we add to it the introduction rule 
$$\irule{}
        {R(\varepsilon)}
        {\mbox{intro}}$$
In particular, if the system contains a neutral rule 
$$\irule{}{R(x)}{\mbox{neutral}}$$
then we add to it the introduction rule 
$$\irule{}{R(\varepsilon)}{\mbox{intro}}$$
\end{itemize}
As there is only a finite number of possible rules, this process 
terminates.
\end{definition}

\begin{example}
\label{ex2}
Consider the system defined in Example \ref{ex}. 
We successively add the following rules 
$$
\begin{array}{llll}
\irule{Q(x)}{S(x)}{\mbox{\bf n3}}
~~~~~~~~~~~~~~
& 
\irule{}{T(\varepsilon)}{\mbox{\bf i5}}
~~~~~~~~~~~~~~
&

\irule{}{T(a x)}{\mbox{\bf i6}}
~~~~~~~~~~~~~~
&
\irule{Q(x)~T(x)}{Q(ax)}{\mbox{\bf i7}}
\\
\\
\irule{Q(x)~T(x)}{S(ax)}{\mbox{\bf i8}}
~~~~~~~~~~~~~~
&
\irule{}{T(b x)}{\mbox{\bf i9}}
~~~~~~~~~~~~~~
&
\irule{T(x)}{Q(bx)}{\mbox{\bf i10}}
~~~~~~~~~~~~~~
&
\irule{T(x)}{S(bx)}{\mbox{\bf i11}}
\end{array}$$
where 
the rule {\bf n3} is obtained from {\bf i1} and {\bf e1}, 
the rule {\bf i5} from {\bf n2}, 
the rule {\bf i6} from {\bf n2},
the rule {\bf i7} from {\bf i1}, {\bf i3}, and {\bf n1}, 
the rule {\bf i8} from {\bf i7} and {\bf n3}, 
the rule {\bf i9} from {\bf n2}, 
the rule {\bf i10} from {\bf i2}, {\bf i4}, and {\bf n1}, and
the rule {\bf i11} from {\bf i10} and {\bf n3}.

Then, no more rules can be added.
\end{example}

\begin{lemma}
If ${\cal I}$ is a small step system, and ${\cal I}_s$ is its saturation, 
then ${\cal I}$ and ${\cal I}_s$ prove the same configurations.
\end{lemma}

\proof{All the rules added in ${\cal I}_s$ are derivable in ${\cal I}$.}

\medskip

Now, we are ready to prove that a saturated system has the 
cut-elimination property.

\begin{lemma}[Cut-elimination]
If a configuration $A$ has a proof $\pi$ in a saturated system, 
it has a cut-free proof.
\end{lemma}

\proof{Assume the proof $\pi$ contains a cut. 
If this cut has the form
$$\irule{\irule{\irule{\pi_1}
                      {P_1(w)}
                      {}
                ~...~
                \irule{\pi_m}
                      {P_m(w)}
                      {}
               }
               {Q_1(\gamma w)}
               {\mbox{intro}}
               ~~~~~~~~~~~~~~~
               \irule{\rho_2}{Q_2(w)}{}~...~
               \irule{\rho_n}{Q_n(w)}{}
        }
        {R(w)}
        {\mbox{elim}}$$
we replace it by the proof 
$$\irule{\irule{\pi_1}{P_1(w)}{}~...~\irule{\pi_m}{P_m(w)}{}~
         \irule{\rho_2}{Q_2(w)}{}~...~\irule{\rho_n}{Q_n(w)}{}}
        {R(w)}
        {\mbox{neutral}}$$
If it has the form
$$\irule{\irule{\irule{\pi^1_1}
                      {P^1_1(w)}
                      {}
                ~...~
                \irule{\pi^1_{m_1}}
                      {P^1_{m_1}(w)}
                      {}
               }
               {Q_1(\gamma w)}
               {\mbox{intro}}
         ~~~~~~~~~~~~~~~~~~~
         ... 
         ~~~~~~~
         \irule{\irule{\pi^n_1}
                      {P^n_1(w)}
                      {}
                ~...~
                \irule{\pi^n_{m_n}}
                      {P^n_{m_n}(w)}
                      {}
                }
                {Q_n(\gamma w)}
               {\mbox{intro}}
        }
        {R(\gamma w)}
        {\mbox{neutral}}
$$
we replace it by the proof 
$$\irule{\irule{\pi^1_1}{P^1_1(w)}{}~...~\irule{\pi^1_{m_1}}{P^1_{m_1}(w)}{}~...~
         \irule{\pi^n_1}{P^n_1(w)}{}~...~\irule{\pi^n_{m_n}}{P^n_{m_n}(w)}{}}
        {R(\gamma w)}
        {\mbox{intro}}$$
If it has the form
$$\irule{\irule{}
               {Q_1(\varepsilon)}
               {\mbox{intro}}
         ~~~~~~~~
         ... 
         ~~
         \irule{}
               {Q_n(\varepsilon)}
               {\mbox{intro}}
        }
        {R(\varepsilon)}
        {\mbox{neutral}}
$$
we replace it by the proof 
$$\irule{}
        {R(\varepsilon)}
        {\mbox{intro}}$$
This process terminates as 
the ordered pair formed with the number of elimination rules and the 
number of neutral rules decreases at each step of the reduction for 
the lexicographic order on ${\mathbb N}^2$.
\begin{example}
In the system of Example \ref{ex2}, the proof
$$\irule{\irule{\irule{\irule{\irule{\irule{\irule{}
                                                  {T(\varepsilon)}
                                                  {\mbox{\bf n2}}
                                           }
                                           {P(b)}
                                           {\mbox{\bf i2}}
                                     ~~~~~~~~~~~~~~~
                                     \irule{}
                                           {R(b)}
                                           {\mbox{\bf i4}}
                                    }
                                    {Q(b)}
                                    {\mbox{\bf n1}}
                             }
                             {P(ab)}
                             {\mbox{\bf i1}}
                      ~~~~~~~~~~~~~~~~~~~~~~~
                      \irule{\irule{}{T(b)}{\mbox{\bf n2}}}
                            {R(ab)}
                            {\mbox{\bf i3}}
                      }
                      {Q(ab)}
                      {\mbox{\bf n1}}
               }
               {P(aab)}
               {\mbox{\bf i1}}
         }
         {S(ab)}
         {\mbox{\bf e1}}$$
reduces to 
$$\irule{\irule{\irule{\irule{\irule{\irule{\irule{}
                                                  {T(\varepsilon)}
                                                  {\mbox{\bf i5}}
                                           }
                                           {P(b)}
                                           {\mbox{\bf i2}}
                                     ~~~~~~~~~~~~~~~
                                     \irule{}
                                           {R(b)}
                                           {\mbox{\bf i4}}
                                    }
                                    {Q(b)}
                                    {\mbox{\bf n1}}
                             }
                             {P(ab)}
                             {\mbox{\bf i1}}
                      ~~~~~~~~~~~~~~~~~~~~~~~
                      \irule{\irule{}{T(b)}{\mbox{\bf n2}}}
                            {R(ab)}
                            {\mbox{\bf i3}}
                      }
                      {Q(ab)}
                      {\mbox{\bf n1}}
               }
               {P(aab)}
               {\mbox{\bf i1}}
         }
         {S(ab)}
         {\mbox{\bf e1}}$$
then to 
$$\irule{\irule{\irule{\irule{\irule{\irule{\irule{}
                                                  {T(\varepsilon)}
                                                  {\mbox{\bf i5}}
                                           }
                                           {P(b)}
                                           {\mbox{\bf i2}}
                                     ~~~~~~~~~~~~~~~
                                     \irule{}
                                           {R(b)}
                                           {\mbox{\bf i4}}
                                    }
                                    {Q(b)}
                                    {\mbox{\bf n1}}
                             }
                             {P(ab)}
                             {\mbox{\bf i1}}
                      ~~~~~~~~~~~~~~~~~~~~~~~
                      \irule{\irule{}{T(b)}{\mbox{\bf i9}}}
                            {R(ab)}
                            {\mbox{\bf i3}}
                      }
                      {Q(ab)}
                      {\mbox{\bf n1}}
               }
               {P(aab)}
               {\mbox{\bf i1}}
         }
         {S(ab)}
         {\mbox{\bf e1}}$$
then to
$$\irule{\irule{\irule{\irule{\irule{\irule{}
                                           {T(\varepsilon)}
                                           {\mbox{\bf i5}}
                                    }
                                    {P(b)}
                                    {\mbox{\bf i2}}
                              ~~~~~~~~~~~~~
                              \irule{}
                                    {R(b)}
                                    {\mbox{\bf i4}}
                             }
                             {Q(b)}
                             {\mbox{\bf n1}}
                      }
                      {P(ab)}
                      {\mbox{\bf i1}}
               ~~~~~~~~~~~~~~~~~~~~~
               \irule{\irule{}{T(b)}{\mbox{\bf i9}}}
                     {R(ab)}
                     {\mbox{\bf i3}}
               }
               {Q(ab)}
               {\mbox{\bf n1}}
        }
        {S(ab)}
        {\mbox{\bf n3}}$$
then to
$$\irule{\irule{\irule{\irule{\irule{}{T(\varepsilon)}{\mbox{\bf i5}}}
                             {Q(b)}
                             {\mbox{\bf i10}}
                      }
                      {P(ab)}
                      {\mbox{\bf i1}}
               ~~~~~~~~~~~~
               \irule{\irule{}{T(b)}{\mbox{\bf i9}}}
                     {R(ab)}
                     {\mbox{\bf i3}}
               }
               {Q(ab)}
               {\mbox{\bf n1}}
        }
        {S(ab)}
        {\mbox{\bf n3}}$$
then to
$$\irule{\irule{\irule{\irule{}{T(\varepsilon)}{\mbox{\bf i5}}}
                      {Q(b)}
                      {\mbox{\bf i10}}
               ~~~~~~~~~~~~
               \irule{}{T(b)}{\mbox{\bf i9}}
               }
               {Q(ab)}
               {\mbox{\bf i7}}
        }
        {S(ab)}
        {\mbox{\bf n3}}$$
and finally to
$$\irule{\irule{\irule{}{T(\varepsilon)}{\mbox{\bf i5}}}
               {Q(b)}
               {\mbox{\bf i10}}
        ~~~~~~~~~~
        \irule{}{T(b)}{\mbox{\bf i9}}
        }
        {S(ab)}
        {\mbox{\bf i8}}$$

\end{example}

\begin{lemma}
A cut-free proof contains introduction rules only.
\end{lemma}

\proof{By induction over proof structure. The proof has the form 
$$\irule{\irule{\pi_1}{A_1}{}~...~\irule{\pi_n}{A_n}{}
        }
        {B}
        {}$$
By induction hypothesis, the proofs $\pi_1$, ..., $\pi_n$ contain introduction 
rules only. As the proof is cut-free, the last rule is neither an elimination
rule, nor a neutral rule. Thus, it is an introduction rule.}

\begin{theorem}
Provability in an alternating pushdown system is decidable.
\end{theorem}

\proof{
If ${\cal I}_0$ is an alternating pushdown system, ${\cal I}$ the small 
step corresponding system, ${\cal I}_s$ its saturation, 
and ${\cal I}'$ the alternating multi-automaton obtained 
by dropping all the elimination rules and all the neutral rules from 
${\cal I}_s$, then ${\cal I}_0$, ${\cal I}$, ${\cal I}_s$, and ${\cal I}'$
prove the same configurations expressed in the language of ${\cal I}_0$
and provability in the alternating multi-automaton ${\cal I}'$ is decidable.}

\medskip

Note that this decidability proof follows the line of \cite{BEM}, in
the sense that, for a given alternating pushdown system, it builds an
alternating multi-automaton recognizing the same configurations.  The
originality of our approach is that, in our setting, alternating
multi-automata are just particular alternating pushdown systems,
while, these concepts are usually defined independently.  This way, we
can avoid building this alternating multi-automaton from
scratch. Rather, we progressively transform the alternating pushdown
system under consideration into an alternating multi-automaton
recognizing the same configurations.

\medskip

As a corollary of the decidability result proved in Section \ref{secdec}, 
we prove that any alternating pushdown system can be extended to a complete
system, where for every configuration $A$, either $A$ or $\neg A$ is provable.
We first recall, in Section \ref{seccoco}, some well-known facts about 
inductive and co-inductive proofs, then we use, in Section \ref{seccoindind}, 
the results of Sections \ref{secdec} and \ref{seccoco} to 
extend alternating pushdown systems to complete systems.

\section{Complementation and co-inductive proofs}
\label{seccoco}

\begin{definition}
\label{D}
An inference system ${\cal I}$ 
defines a function $F_{\cal I}$ mapping a set of configurations $X$ to the
set of configurations that can be deduced in one step with the rules of 
${\cal I}$ from the configurations of $X$:
$$F_{\cal I}(X) = \{\sigma B \in {\cal P}~|~\exists A_1 ... A_n~\mbox{s.t.}~
\sigma A_1 \in X, ..., \sigma A_n \in X,~\mbox{and}~\frac{A_1~...~A_n}{B} 
\in {\cal I}\}$$
where ${\cal P}$ is the set of all configurations.
\end{definition}

It is well-known that the function $F_{\cal I}$ is continuous, that is, 
for all increasing sequences 
$X_0, X_1, ...$ of sets of configurations, 
$F_{\cal I}(\bigcup_n X_n) = \bigcup_n F_{\cal I}(X_n)$. 
Thus, this function $F_{\cal I}$ has a least fixed point 
$$D = \bigcup_n F_{\cal I}^n(\varnothing)$$
and a configuration $A$ is an element of $D$ if and only if it has a proof 
in the sense of Definition \ref{proof}. 

\begin{definition}[Conjugate function]
Consider an inference system ${\cal I}$ and the associated function 
$F_{\cal I}$. The {\em conjugate} $G_{\cal I}$ of the function
$F_{\cal I}$ is defined by 
$$G_{\cal I}(X) = {\cal P} \setminus F_{\cal I}({\cal P} \setminus X)$$
\end{definition}

\begin{lemma}
\label{complement}
Let ${\cal I}$ be an inference system. 
The function $G_{\cal I}$ is co-continuous, that is, for all decreasing 
sequences $X_0, X_1, ...$ of sets of configurations, one has 
$G_{\cal I}(\bigcap_n X_n) = \bigcap_n G_{\cal I}(X_n)$
and the complement of the set $D$, of Definition \ref{D}, is the 
greatest fixed point of this function:
$${\cal P} \setminus D = \bigcap_n G_{\cal I}^n({\cal P})$$
\end{lemma}

\proof{It is easy to check, using the definition of $G_{\cal I}$ and
the continuity of $F_{\cal I}$, that $G_{\cal I}$ is co-continuous.
Then, by induction on $n$, we prove that $G_{\cal I}^n({\cal P}) = {\cal
P} \setminus F_{\cal I}^n(\varnothing)$ and with 
${\cal P} \setminus \bigcup_n F_{\cal I}^n(\varnothing)
= \bigcap_n ({\cal P} \setminus F_{\cal I}^n(\varnothing))$, we conclude that
${\cal P} \setminus D = \bigcap_n G_{\cal I}^n({\cal P})$.}

\medskip

We now focus on inference systems ${\cal I}$, such 
that the function $G_{\cal I}$ can 
be defined with an inference system $\overline{{\cal I}}$, 
the complementation of ${\cal I}$ defined below.

\begin{lemma}
\label{concl}
For each small step alternating pushdown system ${\cal I}$, 
we can build an equivalent inference system
$\tilde{\cal I}$ and a set ${\cal C}$ such that 
\begin{itemize}
\item the conclusions of the rules of $\tilde{\cal I}$ are in ${\cal C}$, 

\item for every configuration $A$ there exists a unique proposition 
$B$ in ${\cal C}$ such that $A$ is an instance of $B$.
\end{itemize}
\end{lemma}

\proof{We take for ${\cal C}$ the set containing all the 
propositions of the form $P(\varepsilon)$ and $P(\gamma x)$. 
Then, we replace each neutral rules and elimination rules with the conclusion 
$P(x)$ by an instance with the conclusion $P(\varepsilon)$ and for each stack 
symbol $\gamma$, an instance with the conclusion $P(\gamma x)$.}

\begin{definition}[Complementation]
Let ${\cal I}$ be a small step alternating pushdown system, 
$\tilde{\cal I}$ the system built at Lemma \ref{concl}, 
and ${\cal C}$ be a finite set of atomic propositions such that 
\begin{itemize}
\item the conclusions of the rules of $\tilde{\cal I}$ are in the set 
${\cal C}$, 

\item for every configuration $A$, there exists a unique proposition 
$B$ in ${\cal C}$ such that $A$ is an instance of $B$.
\end{itemize}
Then, we define the system $\overline{\cal I}$ , the 
{\em complementation} of ${\cal I}$, as follows: 
for each $B$ in ${\cal C}$, if the system $\tilde{\cal I}$ contains
$n$ rules $r^B_1, ..., r^B_n$ with the conclusion $B$, where $n$ may be zero,
$$\irule{A^1_1~...~A^1_{m_1}}
        {B}
        {}$$
$$...$$
$$\irule{A^n_1~...~A^n_{m_n}}
        {B}
        {}$$
then the system $\overline{\cal I}$ contains the $m_1 ... m_n$ rules 
$$\irule{A^1_{j_1}~...~A^n_{j_n}}
        {B}
        {}$$
\end{definition}

\begin{example}
\label{examplecoind}
Consider the language containing a constant $\varepsilon$, a monadic 
function symbol $a$, and monadic predicate symbols $P$, $Q$, $R$, $S$. 
Consider the small step inference system ${\cal R}$ 
$$\begin{array}{llll}
\irule{Q(x)~~~R(x)}
      {P(x)}
      {}
~~~~~~~~~~~~~~~~~~~~~~
&
\irule{S(x)}
      {P(x)}
      {}
~~~~~~~~~~~~~~~~~~~~~~
&
\irule{P(a x)}
      {Q(x)}
      {}
~~~~~~~~~~~~~~~~~~~~~~
&
\irule{}
      {R(a x)}
      {}
\end{array}$$
we transform this system into the equivalent inference system $\tilde{\cal R}$
$$\begin{array}{llll}
\irule{Q(\varepsilon)~~~R(\varepsilon)}
      {P(\varepsilon)}
      {}
~~~~~~~~~~~~~~~~~~~~~~
&
\irule{Q(a x)~~~R(a x)}
      {P(a x)}
      {}

~~~~~~~~~~~~~~~~~~~~~~
&
\irule{S(\varepsilon)}
      {P(\varepsilon)}
      {}
~~~~~~~~~~~~~~~~~~~~~~
&
\irule{S(a x)}
      {P(a x)}
      {}
\\
\\
\irule{P(a)}
      {Q(\varepsilon)}
      {}
~~~~~~~~~~~~~~~~~~~~~~
&
\irule{P(a a x)}
      {Q(a x)}
      {}
~~~~~~~~~~~~~~~~~~~~~~
&
\irule{}
      {R(a x)}
      {}
\end{array}$$
Then, the system $\overline{\cal R}$ is defined by the rules
$$\begin{array}{llll}
\irule{Q(\varepsilon)~~~S(\varepsilon)}
      {P(\varepsilon)}
      {}
~~~~~~~~~~~~~~~~~~~~~~
&
\irule{R(\varepsilon)~~~S(\varepsilon)}
      {P(\varepsilon)}
      {}
~~~~~~~~~~~~~~~~~~~~~~
&
\irule{Q(a x)~~~S(a x)}
      {P(a x)}
      {}
~~~~~~~~~~~~~~~~~~~~~~
&
\irule{R(a x)~~~S(a x)}
      {P(a x)}
      {}
\\
\\
\irule{P(a)}
      {Q(\varepsilon)}
      {}
~~~~~~~~~~~~~~~~~~~~~~
&
\irule{P(aax)}
      {Q(ax)}
      {}
~~~~~~~~~~~~~~~~~~~~~~
&
\irule{}
      {R(\varepsilon)}
      {}
~~~~~~~~~~~~~~~~~~~~~~
&
\irule{}
      {S(\varepsilon)}
      {}
\\
\\
\irule{}
      {S(a x)}
      {}
\\
\\
\end{array}$$
\end{example}

\begin{lemma}
\label{complement2}
The function $F_{\overline{\cal I}}$ is the function $G_{\tilde{\cal I}}$, 
that is, a configuration is provable in $\overline{\cal I}$ in one step from 
the set of configurations
${\cal P} \setminus X$, if and only if it is not provable 
in one step in $\tilde{\cal I}$ from the set of configurations $X$.
\end{lemma}

\proof{Consider a configuration $B$.
There exists a 
unique proposition $C$ in ${\cal C}$ such that 
$B = \sigma C$.

Given a set of configurations X, assume $B$ is provable in one step from 
${\cal P} \setminus X$ 
with a rule of $\overline{\cal I}$, then the premises 
$\sigma A^i_{j_i}$ are in ${\cal P} \setminus X$. 
Thus none of these configurations is in $X$, thus $B$ is not provable
in one step from $X$ with a rule of $\tilde{\cal I}$.

Conversely, assume $B$ is not provable in one step in $\tilde{\cal I}$ 
from the configurations of $X$, then for each 
inference rule with the conclusion $C$, $r^C_i$ of $\tilde{\cal I}$, there
exists a premise $A^i_{j_i}$ such that $\sigma A^i_{j_i}$ is not an element of 
$X$. 
Thus, all the configurations $\sigma A^i_{j_i}$ are in ${\cal P} \setminus X$ 
and hence $B$ is provable in one step from 
${\cal P} \setminus X$ with a rule of $\overline{\cal I}$.}

\begin{definition}[Co-inductive proof]
A {\em co-inductive proof} in an inference system ${\cal J}$ 
is a finite or infinite tree labeled by 
configurations such that for each node $N$, there exists an inference rule 
$$\irule{A_1~...~A_n}{B}{}$$ in ${\cal J}$, 
and a substitution $\sigma$ such that the node $N$ is labeled with 
$\sigma B$ and its children are labeled with
$\sigma A_1$, ..., $\sigma A_n$. 
A co-inductive proof is a {\em co-inductive proof of a configuration} $A$ if 
its root is labeled by $A$.
A configuration $A$ is said to be {\em co-inductively provable} 
if it has a co-inductive proof.
\end{definition}

It is well-known that a configuration $A$ is an element of the greatest
fixed point of the co-continuous function $F_{\cal J}$ if and only if 
it has a co-inductive proof in the system ${\cal J}$ \cite{Sangiorgi}.

\begin{theorem}
Let ${\cal I}$ be a small step alternating pushdown system.
A configuration has a co-inductive proof in $\overline{\cal I}$ if 
and only if it has no proof in ${\cal I}$.
\end{theorem}

\proof{A configuration $A$ has a co-inductive proof in $\overline{\cal I}$
if and only it is an element of the greatest fixed point of the co-continuous 
function $F_{\overline{\cal I}}$, if and only if it is an element of the 
greatest fixed point of the
co-continuous function $G_{\tilde{\cal I}}$ (by Lemma \ref{complement2}), 
if and only if it is not an element of the least fixed point of the function 
$F_{\tilde {\cal I}}$ (by Lemma \ref{complement}), 
if and only if it has no proof in $\tilde{\cal I}$
if and only if it has no proof in ${\cal I}$ (by Lemma \ref{concl}).}

\begin{example}
\label{examplecoind2}
The configuration $P(a)$ is not provable in the system ${\cal R}$
defined in Example \ref{examplecoind}, and
it has a co-inductive proof in the system $\overline{\cal R}$:

$$\irule{\irule{\irule{\irule{\irule{...}
                                    {P(aaa)}
                                    {}
                             }
                             {Q(aa)}
                             {}
                       ~~~~~~~~~
                       \irule{}
                             {S(aa)}
                             {}
                      }
                      {P(aa)}
                      {}
               }
               {Q(a)}
               {}
         ~~~~~~~~~
         \irule{}
               {S(a)}
               {}
        }
        {P(a)}
        {}$$
\end{example}

This result can be used to introduce negation as failure in alternating
pushdown systems. Instead of defining another system $\overline{\cal I}$, we 
just extend the system ${\cal I}$ into a system ${\cal I}_{\neg}$ 
with the rules
$$\irule{\neg A^1_{j_1}~...~\neg A^n_{j_n}}
        {\neg B}
        {}$$
However, this requires to consider co-inductive proofs for closed
propositions of the form $\neg A$ and usual inductive proofs for closed
propositions of the form $A$, as illustrated in Example
\ref{examplecoind2}.

\section{From co-inductive proofs to inductive proofs}
\label{seccoindind}

To avoid to consider co-inductive proofs for closed propositions of the form 
$\neg A$, as we did in Section \ref{seccoco}, we can first transform a
small step
alternating pushdown system
${\cal I}$ into a saturated alternating pushdown system ${\cal I}_s$
and then into an alternating multi-automaton ${\cal I}'$ and then
transform ${\cal I}'$ into ${\cal I}'_{\neg}$ 
\begin{diagram}[height=2em,width=2em]
{\cal I}&\rTo&{\cal I}_s&\rTo&{\cal I}'\\
\dTo&&&&\dTo\\
\tilde{\cal I}&&&&\\
\dTo&&&&\\
{\cal I}_{\neg}&&&&{\cal I}'_{\neg}\\
\end{diagram}
Then, in the rules of system ${\cal I}'_{\neg}$, the premises are always 
smaller than the conclusion. Thus, a co-inductive proof in 
${\cal I}'_{\neg}$ is always finite. This leads to the following theorem.

\begin{theorem}
\label{thfiniteinfinite}
The proposition $\neg A$ has a (finite) proof in ${\cal I}'_{\neg}$ 
if and only if it has a co-inductive proof in ${\cal I}_{\neg}$.
\end{theorem}

\proof{The proposition $\neg A$ has a (finite) proof in ${\cal
I}'_{\neg}$ if and only if it has a co-inductive proof in ${\cal
I}'_{\neg}$ if and only if $A$ has no proof in ${\cal I}'$ if and
only if $A$ has no proof in ${\cal I}$ if and only if $\neg A$ has a
co-inductive proof in ${\cal I}_{\neg}$.}

\begin{example}
As the system ${\cal R}$, defined in Example \ref{examplecoind},
is saturated, a configuration $A$ is provable 
in ${\cal R}$ if and only if it is provable in the system ${\cal R}'$ 
containing only the introduction rule.
$$\irule{}
        {R(a x)}
        {}$$
The system ${\cal R}'_{\neg}$ contains this introduction rule and the 
rules
$$\begin{array}{llll}
\irule{}
      {\neg P(\varepsilon)}
      {}
~~~~~~~~~~~~~~~~~~~~~~
&
\irule{}
      {\neg P(a x)}
      {}
~~~~~~~~~~~~~~~~~~~~~~
&
\irule{}
      {\neg Q(\varepsilon)}
      {}
~~~~~~~~~~~~~~~~~~~~~~
&
\irule{}
      {\neg Q(a x)}
      {}
\\
\\
\irule{}
      {\neg R(\varepsilon)}
      {}
~~~~~~~~~~~~~~~~~~~~~~
&
\irule{}
      {\neg S(\varepsilon)}
      {}
~~~~~~~~~~~~~~~~~~~~~~
&
\irule{}
      {\neg S(a x)}
      {}
\\
\\
\end{array}$$
and the proposition $\neg P(a)$ has the finite proof
$$\irule{}
        {\neg P(a)}
        {}$$
\end{example}

From Theorem \ref{thfiniteinfinite}, if a proposition $\neg A$
has a finite proof in ${\cal I}'_{\neg}$, it has a co-inductive proof in 
${\cal I}_{\neg}$. This result has a more complex, but more informative
proof, where from a finite proof of $\neg A$ in ${\cal I}'_{\neg}$
we reconstruct a co-inductive proof in ${\cal I}_{\neg}$. Such a
co-inductive proof in the complementation of the original system
${\cal I}$ is more informative than the proof in ${\cal I}'_{\neg}$
because it contains an explicit counter-example to $A$: for instance
the proof 

$$\irule{\irule{\irule{\irule{\irule{...}
                                    {\neg P(aaa)}
                                    {}
                             }
                             {\neg Q(aa)}
                             {}
                       ~~~~~~~~~
                       \irule{}
                             {\neg S(aa)}
                             {}
                      }
                      {\neg P(aa)}
                      {}
               }
               {\neg Q(a)}
               {}
         ~~~~~~~~~
         \irule{}
               {\neg S(a)}
               {}
        }
        {\neg P(a)}
        {}$$
explains that $P(a)$ is false because $Q(a)$ and $S(a)$ are 
false, $Q(a)$ is false because $P(aa)$ is false, etc.

\begin{lemma}
\label{combinatorial}
Consider a natural number $n \geq 1$, $n$ families of sets 
$\langle H^1_1, ..., H^1_{k_1} \rangle$, ..., 
$\langle H^n_1, ..., H^n_{k_n} \rangle$ and 
a set $S$, such that each of the $k_1 ... k_n$ sets of the form 
$H^1_{j_1} \cup ... \cup H^n_{j_n}$ contains an element of $S$.
Then, there exists an index $l$, $1 \leq l \leq n$, such that each of 
the sets $H^l_1, ..., H^l_{k_l}$ contains an element of $S$.
\end{lemma}

\proof{By induction on $n$. 

If $n = 1$, then each of the sets $H^1_1$, ..., $H^1_{k_1}$ contains an 
element of $S$.

Then, assume the property holds for $n$ and consider 
$\langle H^1_1, ..., H^1_{k_1} \rangle$, ..., 
$\langle H^n_1, ..., H^n_{k_n} \rangle$, 
$\langle H^{n+1}_1, ..., H^{n+1}_{k_{n+1}} \rangle$ 
such that each of the $k_1 ... k_n k_{n+1}$ sets of the form 
$H^1_{j_1} \cup ... \cup H^n_{j_n} \cup H^{n+1}_{j_{n+1}}$ contains an element of 
$S$.
We have,
\begin{itemize}
\item each of the $k_1 ... k_n$ sets of the form 
$(H^1_{j_1} \cup ... \cup H^n_{j_n}) \cup H^{n+1}_1$ contains an element of 
$S$, 
\item ..., 
\item each of the $k_1 ... k_n$ sets of the form 
$(H^1_{j_1} \cup ... \cup H^n_{j_n}) \cup H^{n+1}_{k_{n+1}}$ contains an element of 
$S$.
\end{itemize}
Thus, 
\begin{itemize}
\item either each of the $k_1 ... k_n$ sets of the form 
$H^1_{j_1} \cup ... \cup H^n_{j_n}$ contains an element of 
$S$ or $H^{n+1}_1$ contains an element of $S$, 
\item ..., 
\item either each of the $k_1 ... k_n$ sets of the form 
$H^1_{j_1} \cup ... \cup H^n_{j_n}$ contains an element of $S$ 
or $H^{n+1}_{k_{n+1}}$ contains an element of $S$.
\end{itemize}
Hence, either 
each of the $k_1 ... k_n$ sets of the form 
$H^1_{j_1} \cup ... \cup H^n_{j_n}$ contains an element of $S$, or 
$H^{n+1}_1$ contains an element of $S$, ..., and $H^{n+1}_{k_{n+1}}$ contains an 
element of $S$. 
Thus, either, by induction hypothesis, there exists an index $l \leq n$ 
such that each of the $H^l_1$, ..., $H^l_{k_l}$ contains an element of $S$,
or each of the sets $H^{n+1}_1$, ..., $H^{n+1}_{k_{n+1}}$ contains an 
element of $S$.
Therefore, there exists an index $l \leq n+1$ such that 
each of the sets $H^l_1$, ..., $H^l_{k_l}$ contains an element of $S$.}

\begin{lemma}
\label{negation}
Let ${\cal I}$ be a small step alternating pushdown system. 
For each rule of ${\cal I}'_{\neg}$ of the form
$$\irule{\neg B_1~...~\neg B_q}
        {\neg A}
        {}$$
there exists a rule of ${\cal I}_{\neg}$ 
$$\irule{\neg C_1~...~\neg C_p}
        {\neg A}
        {}$$
such that the $\neg C_1, ..., \neg C_p$ are provable in ${\cal I}'_{\neg}$ 
from the hypotheses $\neg B_1$, ..., $\neg B_q$.
\end{lemma}

\proof{The rules in ${\cal I}'_{\neg}$ whose conclusion is a negation
have the form
$$\irule{\neg S_1(x)~...~\neg S_q(x)}
        {\neg P(a x)}
        {}$$
and
$$\irule{}
        {\neg P(\varepsilon)}
        {}$$
Consider first a rule of the form
$$\irule{\neg S_1(x)~...~\neg S_q(x)}
        {\neg P(a x)}
        {}$$
By the construction of ${\cal I}_{\neg}$, it is sufficient to prove that
each rule of $\tilde{\cal I}$ with the conclusion $P(a x)$
has a premise whose negation is provable in ${\cal I}'_{\neg}$ from 
the hypotheses $\neg S_1(x)$, ..., $\neg S_q(x)$.

\begin{itemize}
\item
Consider an introduction rule in $\tilde{\cal I}$
$$\irule{Q_1(x)~...~Q_n(x)}
        {P(a x)}
        {}$$
This rule is also a rule of ${\cal I}$, ${\cal I}_s$ and
${\cal I}'$, thus, by construction of 
${\cal I}'_{\neg}$, one of the $S_i(x)$ is a $Q_j(x)$, thus 
$\neg Q_j(x)$ is provable in ${\cal I}'_{\neg}$ from 
$\neg S_1(x)$, ..., $\neg S_q(x)$.

\item
Consider a rule of $\tilde{\cal I}$
$$\irule{Q_1(a x)~...~Q_n(a x)}
        {P(a x)}
        {}$$
instance of a neutral rule of ${\cal I}$
$$\irule{Q_1(x)~...~Q_n(x)}
        {P(x)}
        {}$$
As there is a rule ${\cal I}'_{\neg}$, with the conclusion 
$\neg P(a x)$,
the number $n$ of premises is at least $1$.
Consider the $k_1$ introduction rules of ${\cal I}_s$ 
with the conclusion $Q_1(a x)$ and respective sets of 
premises $H^1_1$, ..., $H^1_{k_1}$, 
..., the $k_n$ introduction rules of ${\cal I}_s$ with the conclusion $Q_n(a x)$
and respective sets of premises $H^n_1$, ..., $H^n_{k_n}$. 
As the system ${\cal I}_s$ is saturated it contains $k_1 ... k_n$ introduction rules 
with the conclusion $P(a x)$ and sets 
of premises of the form $H^1_{j_1} \cup ... \cup H^n_{j_n}$. 
All these rules
are rules of ${\cal I}'$ thus, by the construction of ${\cal I}'_{\neg}$, 
each of these $k_1 ... k_n$ sets contains an element of 
$\{S_1(x), ..., S_q(x)\}$. Thus, 
by Lemma \ref{combinatorial}, there exists an index $l$ 
such that each $H^l_j$ contains an element of $\{S_1(x), ..., S_q(x)\}$.
Thus, by construction, the system ${\cal I}'_{\neg}$
contains a rule deducing the proposition 
$\neg Q_l(a x)$
from premises in 
$\{\neg S_1(x), ..., \neg S_q(x)\}$ and thus
$\neg Q_l(a x)$ is provable in ${\cal I}'_{\neg}$ from 
$\neg S_1(x), ..., \neg S_q(x)$.

\item 
Consider a rule of $\tilde{\cal I}$
$$\irule{Q_1(b a x)~Q_2(a x)~...~Q_n(a x)}
        {P(a x)}
        {}$$
instance of an elimination rule of ${\cal I}$
$$\irule{Q_1(b x)~Q_2(x)~...~Q_n(x)}
        {P(x)}
        {}$$
Consider the $k$ introduction rules of ${\cal I}_s$ 
with the conclusion $Q_1(b x)$ and respective sets of 
premises $H_1$, ..., $H_k$.
As the system ${\cal I}_s$ is saturated it contains $k$ neutral rules 
with the conclusion $P(x)$ and sets 
of premises of the form $H_j \cup \{Q_2(x), ..., Q_n(x)\}$. 
Consider the instances of these neutral rules 
with the conclusion $P(a x)$ and premises 
$(ax/x)H_j \cup \{Q_2(a x), ..., Q_n(a x)\}$. 
By the previous case, each of these $k$ sets
contains an element whose negation is 
provable in ${\cal I}'_{\neg}$ from $\neg S_1(x), ..., \neg S_q(x)$.
Thus, either one of the $\neg Q_i(a x)$ 
is provable in ${\cal I}'_{\neg}$ from 
$\neg S_1(x), ..., \neg S_q(x)$, 
or each of the sets $(ax/x)H_1$, ..., $(ax/x)H_k$ contains 
an element whose negation is 
provable in ${\cal I}'_{\neg}$ from
$\neg S_1(x), ..., \neg S_q(x)$
in which case $\neg Q_1(b a x)$ is provable 
in ${\cal I}'_{\neg}$ from $\neg S_1(x), ..., \neg S_q(x)$.
\end{itemize}
The proof is similar for rules of the form
$$\irule{}
        {\neg P(\varepsilon)}
        {}$$
By the construction of ${\cal I}_{\neg}$, it is sufficient to prove that
each rule of $\tilde{\cal I}$ with the conclusion $P(\varepsilon)$
has a premise whose negation is provable in ${\cal I}'_{\neg}$.
\begin{itemize}
\item
As ${\cal I}'_{\neg}$ contains the rule
$$\irule{}{\neg P(\varepsilon)}{}$$
there is no rule in ${\cal I}'$ with the conclusion 
$P(\varepsilon)$.
Thus, there is no introduction rule, 
in ${\cal I}_s$, in ${\cal I}$, hence in $\tilde{\cal I}$, 
with the conclusion $P(\varepsilon)$.

\item
Consider a rule of $\tilde{\cal I}$
$$\irule{Q_1(\varepsilon)~...~Q_n(\varepsilon)}
        {P(\varepsilon)}
        {}$$
instance of a neutral rule of ${\cal I}$
$$\irule{Q_1(x)~...~Q_n(x)}
        {P(x)}
        {}$$
As there is a rule ${\cal I}'_{\neg}$, with the conclusion 
$\neg P(\varepsilon)$, the number $n$ of premises is at least $1$.
As the system ${\cal I}_s$ is saturated and contains no
introduction rule with the conclusion 
$P(\varepsilon)$, there exists an index $i$ such that 
there is no introduction rule in ${\cal I}_s$
of the form 
$$\irule{}{Q_i(\varepsilon)}{}$$
Hence, there is no such introduction rule in ${\cal I}'$.
Thus, the system ${\cal I}'_{\neg}$, contains the rule 
$$\irule{}{\neg Q_i(\varepsilon)}{}$$
and the proposition $\neg Q_i(\varepsilon)$ is provable in 
${\cal I}'_{\neg}$. 

\item 
Consider a rule of $\tilde{\cal I}$
$$\irule{Q_1(b)~Q_2(\varepsilon)~...~Q_n(\varepsilon)}
        {P(\varepsilon)}
        {}$$
instance of an elimination rule of ${\cal I}$
$$\irule{Q_1(b x)~Q_2(x)~...~Q_n(x)}
        {P(x)}
        {}$$
Consider the $k$ introduction rules of ${\cal I}_s$ 
with the conclusion $Q_1(b x)$ and respective sets of 
premises $H_1$, ..., $H_k$.
As the system ${\cal I}_s$ is saturated it contains $k$ neutral rules 
with the conclusion $P(x)$ and sets 
of premises of the form $H_j \cup \{Q_2(x), ..., Q_n(x)\}$. 
Consider the instances of these neutral rules 
with the conclusion $P(\varepsilon)$ and premises 
$(\varepsilon/x)H_j \cup \{Q_2(\varepsilon), ..., Q_n(\varepsilon)\}$. 
By the previous case, each of these $k$ sets
contains an element whose negation is 
provable in ${\cal I}'_{\neg}$. 
Thus either one of the $\neg Q_i(\varepsilon)$ 
is provable in ${\cal I}'_{\neg}$
or each of the sets $(\varepsilon/x)H_1$, ..., $(\varepsilon/x)H_k$ contains 
an element whose negation is 
provable in ${\cal I}'_{\neg}$ 
in which case $\neg Q_1(b)$ is provable 
in ${\cal I}'_{\neg}$.
\end{itemize}}

\begin{example}
In the system of Example \ref{examplecoind}, consider
the rule of ${\cal R}'_{\neg}$
$$\irule{}
        {\neg P(a x)}
        {}$$
Both rules of $\tilde{\cal R}$
$$\irule{Q(a x)~~~R(a x)}
        {P(a x)}
        {}$$
and 
$$\irule{S(a x)}
        {P(a x)}
        {}$$
have a premise whose negation is provable in 
${\cal R}'_{\neg}$: $Q(a x)$ for the first and $S(a x)$ for the second.
Thus the rule of ${\cal R}_{\neg}$
$$\irule{\neg Q(a x)~~~\neg S(a x)}
        {\neg P(a x)}
        {}$$
deduces $\neg P(a x)$ from 
premises $\neg Q(a x)$ and $\neg S(a x)$ that are both provable in 
${\cal R}'_{\neg}$. 

In the same way, the system ${\cal R}'_{\neg}$ contains the 
rule
$$\irule{}
        {\neg Q(a x)}
        {}$$
and the rule of ${\cal R}_{\neg}$
$$\irule{\neg P(a a x)}
        {\neg Q(a x)}
        {}$$
deduces $\neg Q(a x)$ from the premise 
$\neg P(a a x)$ that is provable in ${\cal R}'_{\neg}$. 

Finally, the system ${\cal R}'_{\neg}$ contains the rule
$$\irule{}
        {\neg S(a x)}
        {}$$
and the rule of ${\cal R}_{\neg}$
$$\irule{}
        {\neg S(a x)}
        {}$$
deduces $\neg S(a x)$ from no premises.
\end{example}

\begin{lemma}
\label{toto}
If the proposition $\neg A$ is provable in 
${\cal I}'_{\neg}$, then there exists a rule in 
${\cal I}_{\neg}$, deducing $\neg A$ from premises that are
all provable in ${\cal I}'_{\neg}$. 
\end{lemma}

\proof{If the last rule of the proof of $\neg A$ has the form 
$$\irule{\neg S_1(x)~...~\neg S_q(x)}
        {\neg P(a x)}
        {}$$
then $A = P(a w)$, and the propositions 
$\neg S_1(w)$, ..., $\neg S_q(w)$ have proofs in 
${\cal I}'_{\neg}$. By Lemma \ref{negation}, there exists a 
rule in ${\cal I}_{\neg}$ deducing $\neg P(a x)$ from premises
that are all provable in 
${\cal I}'_{\neg}$ from $\neg S_1(x)$, ..., $\neg S_q(x)$. 
Thus this rule deduces $\neg P(a w)$ from premises that are 
provable in ${\cal I}'_{\neg}$ from 
$\neg S_1(w)$, ..., $\neg S_q(w)$. As these propositions are
provable in ${\cal I}'_{\neg}$, so are the premises.

If the last rule of the proof of $\neg A$ has the form 
$$\irule{}
        {\neg P(\varepsilon)}
        {}$$
then $A = P(\varepsilon)$. 
By Lemma \ref{negation}, there exists a 
rule in ${\cal I}_{\neg}$ deducing $\neg P(\varepsilon)$ from premises
that are all provable in 
${\cal I}'_{\neg}$.}

\begin{theorem}
If a proposition $\neg A$ has a proof in the system
${\cal I}'_{\neg}$, then it has a co-inductive proof in the system.
${\cal I}_{\neg}$. 
\end{theorem}

\proof{By Lemma \ref{toto}, the proposition $\neg A$ can be proved
with a rule of ${\cal I}_{\neg}$ whose premises are provable in
${\cal I}'_{\neg}$.  We co-inductively build a proof of these premises.}

\begin{example}
In the system of Example \ref{examplecoind}, consider
the proof in ${\cal R}'_{\neg}$
$$\irule{}{\neg P(a)}{}$$
This proof can be transformed into the proof in ${\cal R}_{\neg}$
$$\irule{\neg Q(a)~~~\neg S(a)}
        {\neg P(a)}
        {}$$
and the proofs in ${\cal R}'_{\neg}$
$$\irule{}{\neg Q(a)}{}$$
and 
$$\irule{}{\neg S(a)}{}$$ 
Applying the same procedure to these premises yields the proof
in ${\cal R}_{\neg}$
$$\irule{\irule{\neg P(aa)}
               {\neg Q(a)}
               {}
         ~~~~~~~~~
         \irule{}
               {\neg S(a)}
               {}
        }
        {\neg P(a)}
        {}$$
and the proof in ${\cal R}'_{\neg}$ 
$$\irule{}{\neg P(aa)}{}$$ 
And iterating this process yields the co-inductive proof in ${\cal R}_{\neg}$
$$\irule{\irule{\irule{\irule{\irule{...}
                                    {\neg P(aaa)}
                                    {}
                             }
                             {\neg Q(aa)}
                             {}
                       ~~~~~~~~~
                       \irule{}
                             {\neg S(aa)}
                             {}
                      }
                      {\neg P(aa)}
                      {}
               }
               {\neg Q(a)}
               {}
         ~~~~~~~~~
         \irule{}
               {\neg S(a)}
               {}
        }
        {\neg P(a)}
        {}$$
\end{example}

\section*{Acknowledgement}

The authors want to thank Ahmed Bouajjani for
enlightening discussions.
This work is supported by the ANR-NSFC project LOCALI (NSFC 61161130530
and ANR 11 IS02 002 01) and the 
Chinese National Basic Research Program (973) Grant No. 2014CB340302.

\end{document}